\documentstyle[12pt,psfig,epsfig]{article}
\def\aprge{\buildrel > \over {_{\sim}}}   
\def\nubar{\overline {\nu} }
 
\begin{document}     
\title{\bf A 3--Dimensional Calculation of \\ 
Atmospheric Neutrino Flux}
\author{G.~Battistoni$^a$, A.~Ferrari$^{a,b}$, P.~Lipari$^c$, T.~Montaruli$^d$ 
\\ 
P.~R.~Sala$^{a,b}$ and T.~Rancati$^a$ } 
\date{}
\maketitle
 
\begin{center}
$a)$ INFN and Dipartimento di Fisica dell' Universit\'a, 20133 Milano, Italy \\
$b)$ now at CERN, Geneva, Switzerland \\
$c)$ INFN and Dipartimento di Fisica dell' Universit\'a ''La Sapienza", 00196 
Roma, Italy \\
$d)$ INFN and Dipartimento di Fisica dell' Universit\'a, 70126 Bari, Italy  \\
\date{}
\maketitle
 
\end{center}
 
\begin{abstract}
 
An extensive 3-dimensional Monte Carlo calculation of the 
atmospheric neutrino flux is in progress with the FLUKA Monte Carlo code.
The results are compared to those obtained under the 1-dimensional approximation,
where secondary particles and decay products are assumed to be collinear to
the primary cosmic ray,
as usually done in most of the already existing flux calculations.
It is shown that the collinear approximation gives rise to a wrong
angular distribution of neutrinos, essentially in the Sub-GeV region.
However, the angular smearing introduced by the experimental
inability of detecting recoils in neutrino interactions with nuclei
is large enough 
to wash out, in practice, most of the 
differences between 3--dimensional and 1--dimensional flux calculations.
Therefore, the use  of the collinear approximation should have not
introduced a significant bias in the determination
of the flavor oscillation parameters in current experiments.
 
\end{abstract}
 
\section{Introduction}
 
The recent experimental observations concerning atmospheric neutrinos
in SuperKamiokande \cite{sk}, Soudan2 \cite{soudan2} and MACRO \cite{macro},
have given new strength to the hypothesis of the existence of
non zero neutrino masses and of flavor neutrino mixing.
 
The precise determination of the oscillation mechanism and of
the oscillation parameters is, and will be in the
next years, still object of intensive work.
Even in view of 
future experimental activities with long--baseline neutrino beams
from accelerators,  atmospheric neutrino experiments are going to play
a fundamental role.
In all cases, the optimization of the detectors must be accompanied
by an improvement of the precision of theoretical prediction both for
the fluxes and for neutrino interactions.
Indeed, the
determination
of the allowed and excluded regions in the oscillation parameter space
for all experiments are heavily relying
on the comparison of experimental data to Monte Carlo predictions.
In this respect, one of the most important inputs is
the deviation of the zenith angle distribution of upward going
$\nu_\mu$'s with respect to the expectation, as observed by SuperKamiokande.
In particular, one of the main new experimental findings of SuperKamiokande, when compared to
the previous experiments, 
is the observation of an angular modulation of both the Sub-GeV 
and Multi-GeV muon neutrino (and antineutrino) samples, while in the progenitor 
detectors like  
Kamiokande \cite{kamioka} or IMB \cite{imb}, apart from considerations concerning
the statistical significance, only the
Multi-GeV sample deviated from expectation\footnote{It has to be reminded that the
due to differences in the containment efficiency, the energy range of Multi-GeV
events is different between SuperKamiokande and the smaller 
Cherenkov detectors.}.
For both the experiments at Kamioka, the expectation is calculated using the simulation 
by Honda et al. \cite{honda}.
Other experiments, like Soudan2 and MACRO, mostly refer to the predictions
based on the Bartol Monte Carlo \cite{bartol}.
 
These two simulations have been performed
by means of different shower codes with different 
hadronic interaction models and different input spectra for the primary cosmic
rays. The importance
of these two aspects for the evaluation of the systematic uncertainty of
neutrino fluxes is still a matter of debate.
Here, we want to focus the discussion on another aspect of existing calculations
which in principle might be relevant for the angular distribution at low 
energies, 
and therefore for the determination 
of the value of oscillation parameters.
As a matter of fact, a common feature of both the Bartol and Honda simulations
(and in those of other authors), is the assumption that the neutrino fluxes 
can be
calculated using a 1--dimensional approach. 
This approximation in practice implies that the
transverse momentum of secondary particles after an interaction or decay 
can be neglected, so that all secondary mesons and eventually their decay
 products,
neutrinos included, are  collinear with the parent cosmic ray particles.
The argument invoked to justify this approach is that 
all possible effects deriving from the choice
of this approximation, if any, are small and are compensated,
in average, by the fact that primary cosmic rays arrive 
from every direction, and therefore a 3-dimensional calculation
can contribute only to establish second order effects, not essential
for the understanding of the relevant physics underneath.
The average angle between
the direction of a neutrino from pion decay, for instance, and the
direction of the primary is expected to be:
\begin{equation}
<\theta_{N\nu}> \simeq \frac{<P^\pi_\perp>}{P_\pi} 
\simeq \frac{0.3~GeV}{4 E_\nu}
=\frac{4.3^\circ}{E_\nu (GeV)}
\label{eq:av}
\end{equation}
The $\pi \nu$ angle can be considered a minor correction.
In the case of $\nu$'s from $\mu$ decay, an additional contribution comes
from the deviation of $\mu^\pm$ in the geomagnetic field. 
We expect this to be small, and, in first approximation, independent of energy,
since the flight path of low energy muons and their curvature radius 
identically scale with $P_\mu$.
 
In view of the improvement in the quality and precision of predictions for
future generation experiments, as Icarus at Gran Sasso \cite{icarus}, a new
full 3-Dimensional calculation, including also the spherical representation
of the atmosphere and of the Earth, has been started. It is based on the
FLUKA Monte Carlo \cite{fluka}, which is a highly detailed and precise code
for the study of transport and interaction of particles in matter, and in
particular for e.m. and hadronic shower simulations.  The work is still in
progress.
Preliminary results were already presented in \cite{taup}. Here
we want to show that
important differences exist, in principle, in the angular distribution of
low energy neutrinos when 3--Dimensional and collinear calculations are
compared.  The collinear approach is justified only above a certain energy
range.  This is a conclusion which holds irrespectively of all the other main
ingredients of shower simulation: primary spectra, geomagnetic field,
hadronic interactions,etc.  The impact on the physics analysis
of neglecting the 3--Dimensional effects depend 
also on the physics of neutrino interactions and on the resolution of a
specific detector in the reconstruction of neutrino direction.
 
In the following, we shall first summarize the main features of
the simulation set up, then we shall discuss
the simulation results. 
These will be presented comparing
the angular distribution in the full 3--D approach to that obtained
in  the collinear
one. We remark that the results presented here do not yet 
include a discussion of the absolute values of neutrino fluxes
coming from this new calculation. 
As a further point, 
we shall consider the features of neutrino charged current 
interactions in order to understand the smearing effect introduced.
We shall attempt some considerations about the relevance of
the 3--D effects for high resolution experiments.
The questions concerning possible differences due to 
the combined effect of geomagnetic cutoff are also briefly reviewed.
The qualitative differences in the results between the 3--D and 1--D approach can
be understood using an analytical approach: this is shown in a final Appendix.

\section{The Simulation Set--up}
 
The FLUKA code \cite{fluka} is a general purpose Monte Carlo code
for the interaction and transport of particles.
It is built and maintained with the aim of including the best possible 
physical models in terms of completeness and precision.
Therefore it contains very detailed
models of electromagnetic, hadron--hadron and hadron--nucleus interactions, 
covering the range from the MeV scale to the many--TeV one. 
Extensive benchmarking against experimental data has been produced (see the 
references
in  \cite{fluka}).
In view of applications for cosmic ray physics, we have implemented a
3--Dimensional spherical representation of the whole Earth 
and of the surrounding atmosphere. 
This is described by a medium composed
by a proper mixture of N, O and Ar, arranged in 51 concentric
spherical shells, each one having a density scaling according to the
known profile of ``standard atmosphere" \cite{atmo}.
Primary particles, sampled
from a continuous spectrum, are injected
at the top of the atmosphere, 
at about 100 km of altitude. The primary flux is assumed to be 
uniform and isotropic.
The flux of all possible secondary products are
scored at different heights in the atmosphere 
and at the Earth boundary. 
Therefore,
we are able to get, besides neutrino fluxes,
the flux of muons and hadrons to be used for the benchmarking against
existing experimental data.
All relevant physics, such as energy loss, transport in magnetic field, 
polarized decay, etc are included in the code by default, and
different geomagnetic models
can be considered. As mentioned before, 
the discussion on the full calculation of neutrino fluxes would need an 
accurate examination of 
all the mentioned ingredients of the code. All this is postponed
to another paper, since this is not relevant for the purpose of this work.
We limit ourselves to address the reader to the quoted references, and to
mention that  we made use of the same primary all--nucleon spectrum used 
by the Bartol group, as in \cite{agrawal}.
The geomagnetic cutoff is applied a posteriori.
Therefore in the first stage the calculation exploits
 the spherical symmetry of the geometry and of the primary flux:
 all points of the sphere surface are equivalent
and can be used to score the neutrino flux. 
This symmetry, in realistic conditions, is broken mostly by the effects of
geomagnetic field. 
In the following, for
sake of clarity, we shall
discuss mainly the angular distribution for the symmetric flux. Further effects
after the inclusion of geomagnetic cutoff are mentioned in the last section.
 
\begin{figure}[htbp]
\begin{center}
\begin{tabular}{c}
\mbox{\epsfig{file=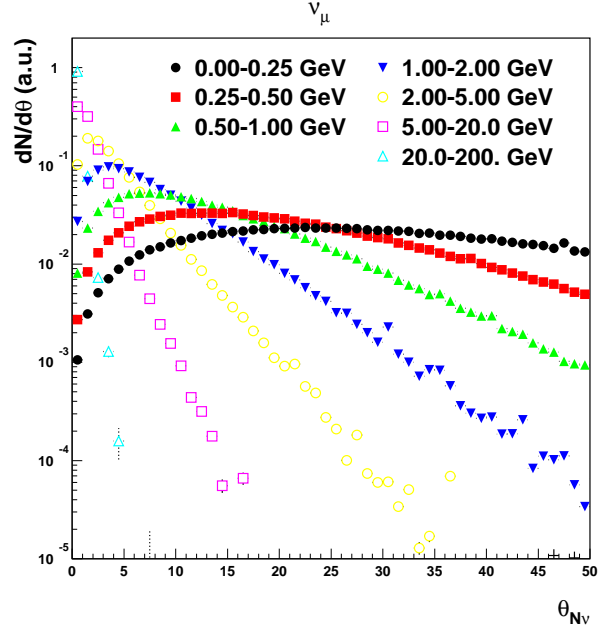,height=9.cm,width=9.cm}} \\
\mbox{\epsfig{file=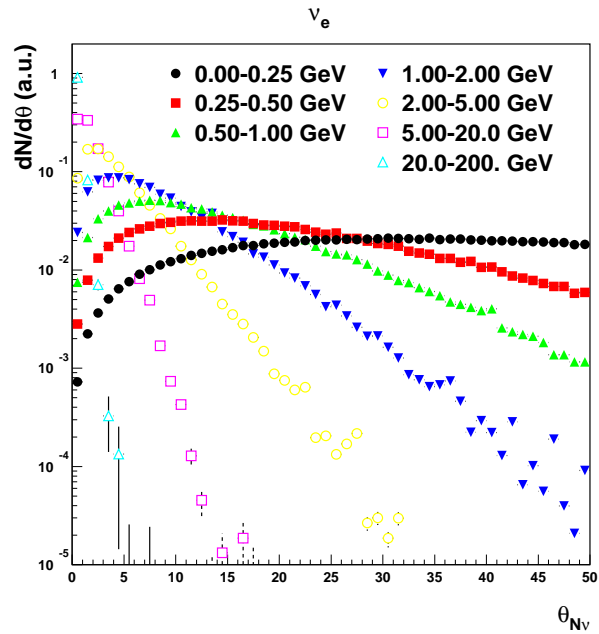,height=9.cm,width=9.cm}}
\end{tabular}
\caption{\em
Calculated distribution of $\theta_{N\nu}$ (degrees) in
different neutrino energy bins.
Up: muon neutrinos.  Down: electron neutrinos.                                         
\label{f:nuang}}
\end{center}
\end{figure}

We can introduce the main topic of this work
by showing a first output from the full simulation, 
that is the distribution of the angle between
a neutrino crossing the Earth's surface and the primary nucleon direction, as
that of Fig. \ref{f:nuang} a) and b) for muon and electron neutrino 
respectively, 
where different intervals of neutrino energy are considered separately.
These distributions have been obtained integrating on the whole energy spectrum
and angular distribution of primaries.
 
From these plots, and from the average values reported in Table
\ref{tb:nuang}, we immediately notice the following features:
\begin{enumerate}
\item While in the Multi-GeV region that
the collinear approximation is reasonable,
in the Sub-GeV region the average angle
with respect to the primary direction is quite large, in substantial
agreement with the expectation from eq.\ref{eq:av}
\item At very low energies,
the average $\theta_{N\nu}$ is somewhat 
larger for electron neutrinos, as also summarized in Table \ref{tb:nuang}. 
This is due to the fact that,
in average, electron neutrinos come from a later stage of the decay chain.
\end{enumerate}
 
\begin{table}[hbt]
\begin{center}
\begin{tabular}{l|rr}
\hline
Energy range (GeV) & $<\theta_{N\nu}>$ (degrees)&\\
 & $\nu_\mu + \nubar_\mu$ & $\nu_e + \nubar_e$ \\
\hline
0.0  - 0.25 & 47.6 & 53.4 \\
0.25 - 0.5  & 23.8 & 27.6 \\
0.5  - 1.0  & 15.6 & 15.9 \\
1.0  - 2.0  &  8.9 &  9.0 \\
2.0  - 5.0  &  4.4 &  4.6 \\
5.0 - 20.0  &  1.8 &  1.8 \\
20.0 - 200.0&  0.5 &  0.5  \\
\hline
\end{tabular}
\caption{\em Average $\theta_{N\nu}$ (angle with respect to primary
particle) for different neutrino flavors and energy intervals.\label{tb:nuang}}
\end{center}
\end{table}
 
\section{Simulation Results}
 
We have performed two different kinds of simulation runs.
The first one exploits all the features of the simulation code.
In the second run we have repeated
the simulation forcing the collinear approach. At each interaction
or decay, the angle of secondary particles with respect to the primary
direction
has been set to the
null value. Furthermore, following a recipe in use in the Bartol 
calculations \cite{gaisser}, all (low energy) neutrinos emitted 
at more than 90$^\circ$ with respect to
the primary direction in the laboratory frame have been disregarded.
 
Before showing the results of numerical simulations,
we want to remark the existence of an effect on the
flux normalization as a function of angle and energy.
This can be understood following the analytical considerations
reported in the Appendix. We also notice that for 
trajectories close to the 
horizontal direction with respect to a given position on the
sphere, there will be cases in which, if perfect collinearity is
assumed, no neutrino will ever touch the Earth surface but
they will all escape through the atmosphere. 
In the 3--Dimensional case, thanks
to a possible large $\theta_{N\nu}$, some neutrinos can instead
be intercepted. 
 
\subsection{Neutrino Flux and Angular Distribution}
 
\begin{figure}[htbp]
\begin{center}
\mbox{\epsfig{file=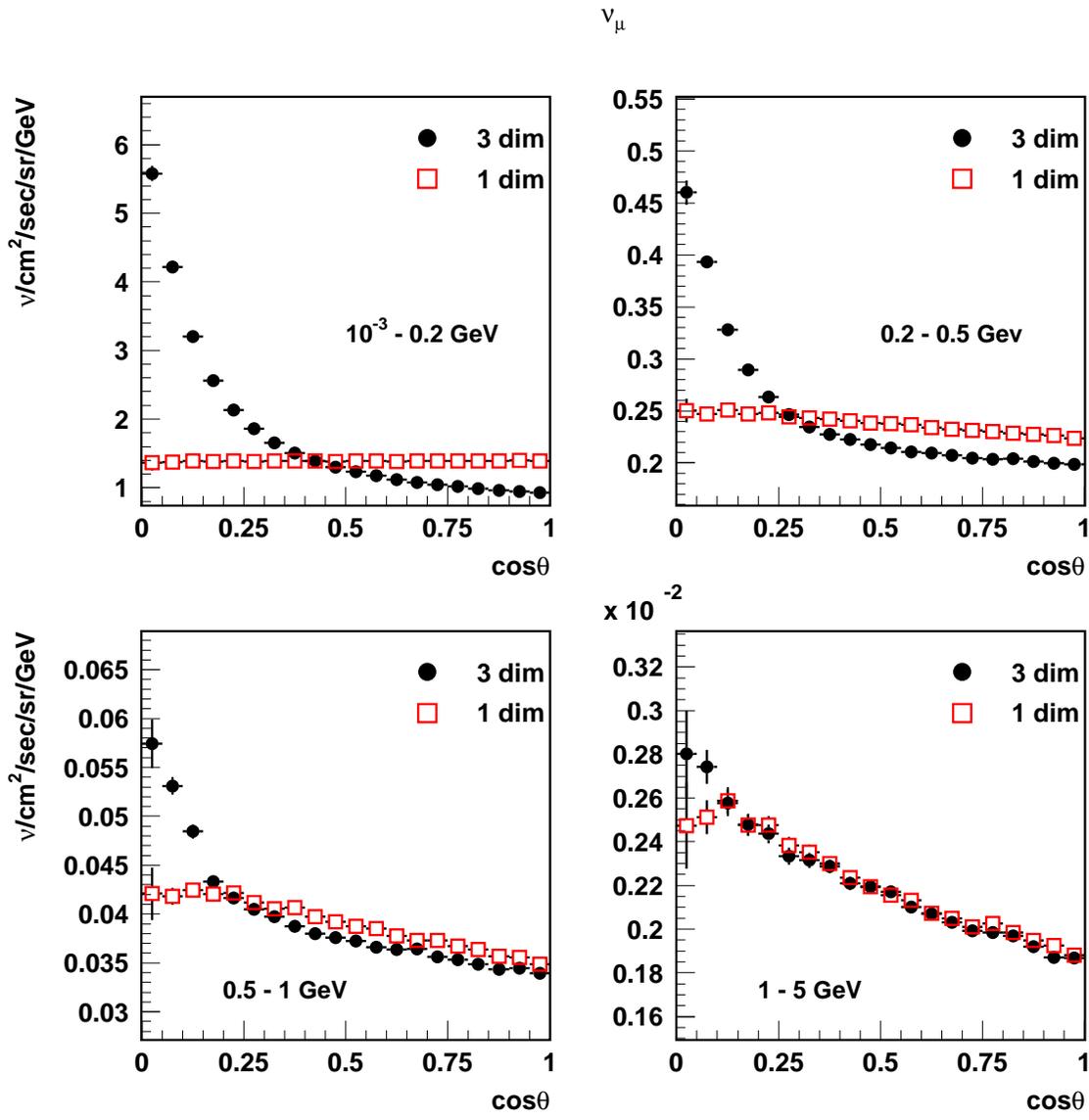,height=15.cm}}
\caption{\em Comparison of the $cos(Zenith)$ dependence 
of $\nu_\mu$ flux in 4 different
energy regions, for the 3-Dimensional and the collinear (1-D) approach,
as obtained in the FLUKA simulation with spherical geometry. \label{f:result1}}
\end{center}
\end{figure}

\begin{figure}[htbp]
\begin{center}
\mbox{\epsfig{file=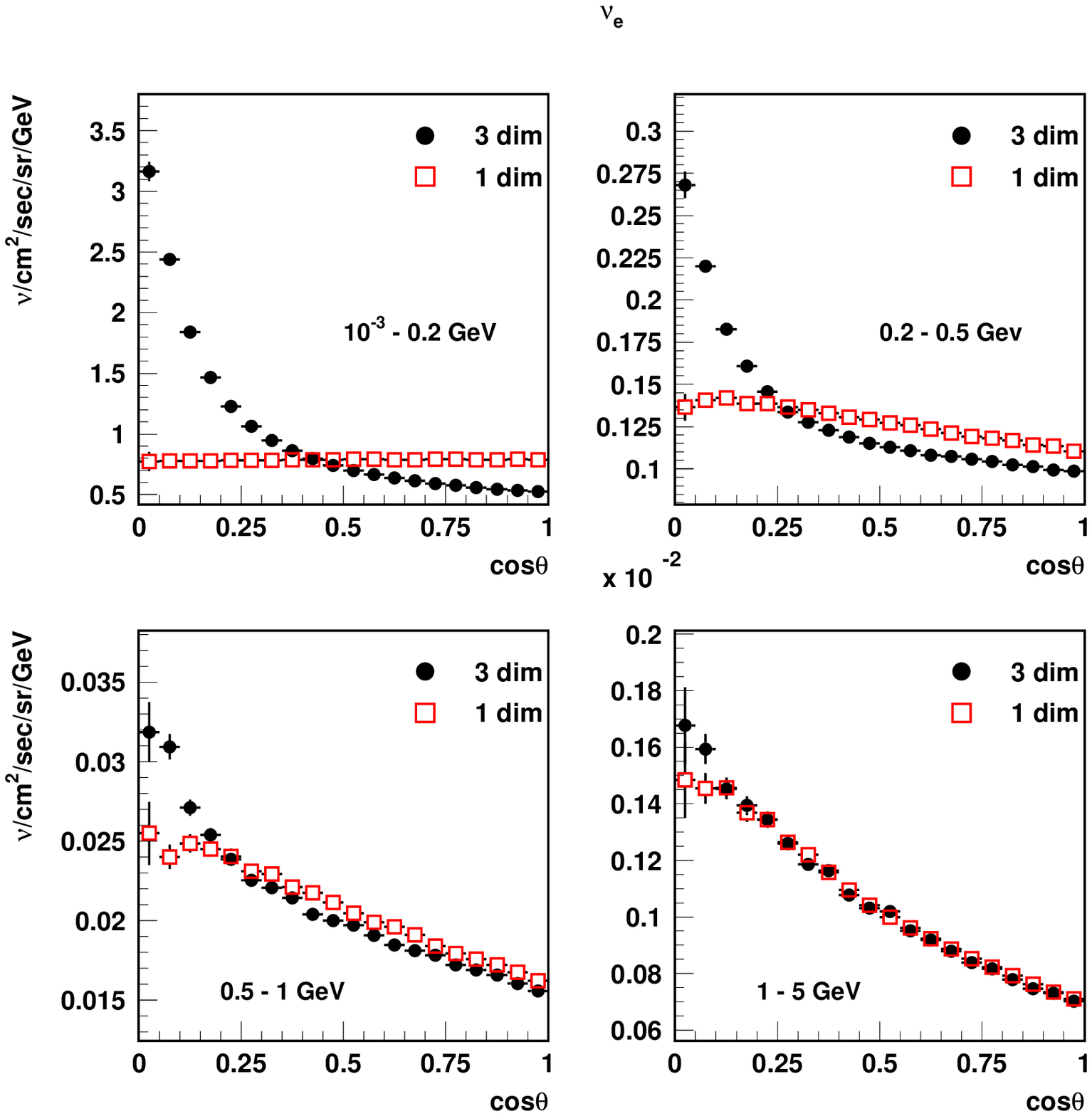,height=15.cm}}
\caption{\em Comparison of $cos(Zenith)$ dependence
of $\nu_e$ flux in 4 different
energy regions, for the 3-Dimensional and the collinear (1-D) approach,
as obtained in the FLUKA simulation with spherical geometry. \label{f:result2}}
\end{center}
\end{figure}

Fig.\ref{f:result1} shows the $cos\theta$ distribution
($\theta$ is the zenith angle) of $\nu_\mu$ for 4 different
energy regions, for the 3-Dimensional and the collinear approach.
Fig. \ref{f:result2} is the analogous plot for $\nu_e$.
Similar plots are obtained for anti--neutrinos.
These results confirm that in the Sub--GeV range, when the average
$\theta_{N\nu}$ is large, there are important differences between the two
approaches. 
The 3-Dimensional calculation predicts an anisotropic angular distribution,
 with an enhancement 
in the horizontal direction, as qualitatively expected by the analytical arguments
which are reported in the Appendix. Conversely, 
the 1-D results give a substantially flat distribution.
We remind that the geomagnetic cut--off has not yet been applied to
the calculated fluxes and perfect spherical symmetry holds.
 
\begin{figure}[htbp]
\begin{center}
\mbox{\epsfig{file=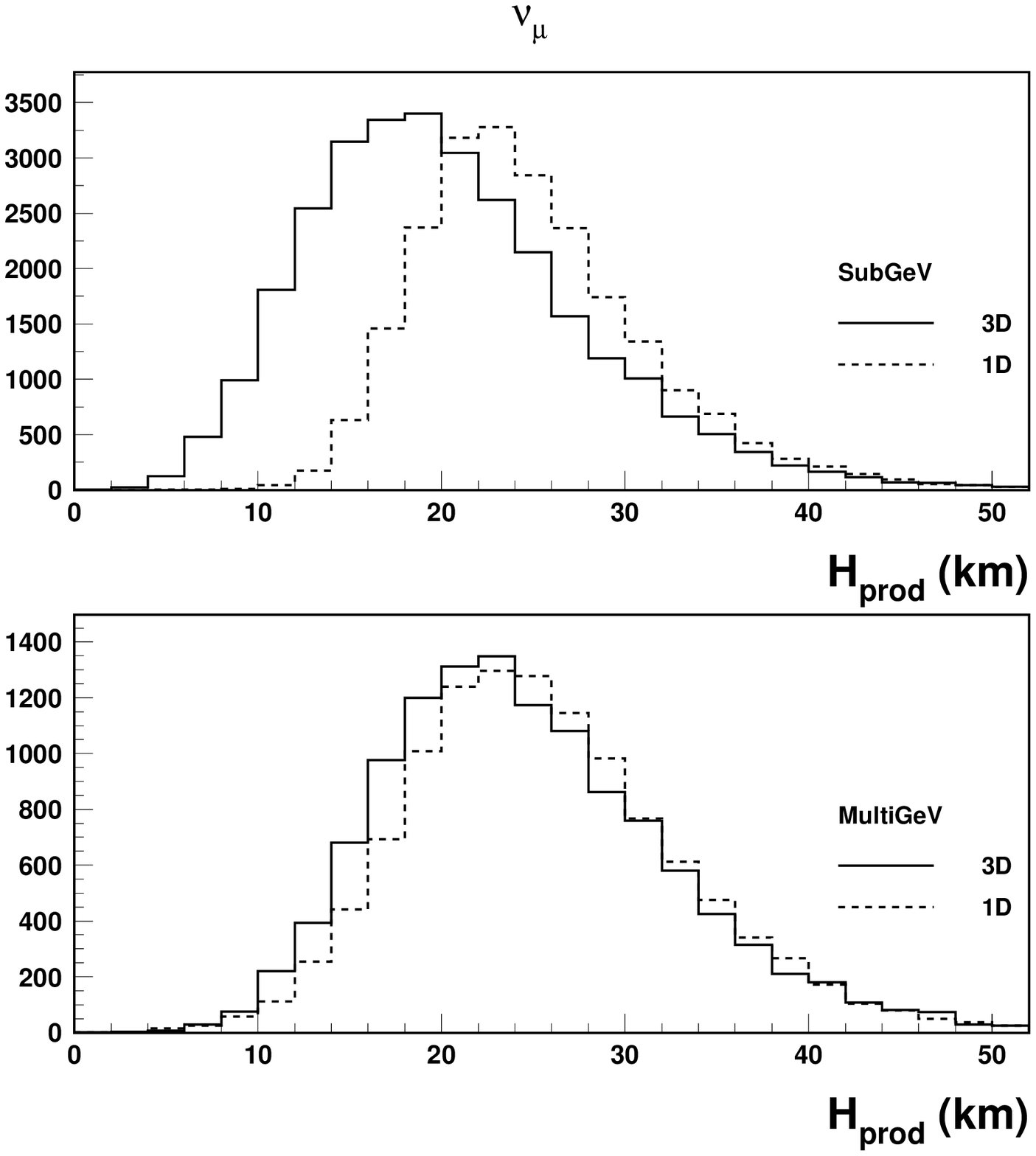,height=15.cm}} 
\caption{\em Distribution of production height of Sub--GeV (up) and Multi--GeV
$\nu_\mu$ and $\nubar_\nu$ around the horizontal ($|cos\theta|<0.2$) as obtained with FLUKA, comparing 3--D
and 1--D simulations.
\label{f:result3}}
\end{center}
\end{figure}

\begin{figure}[htbp]
\begin{center}
\mbox{\epsfig{file=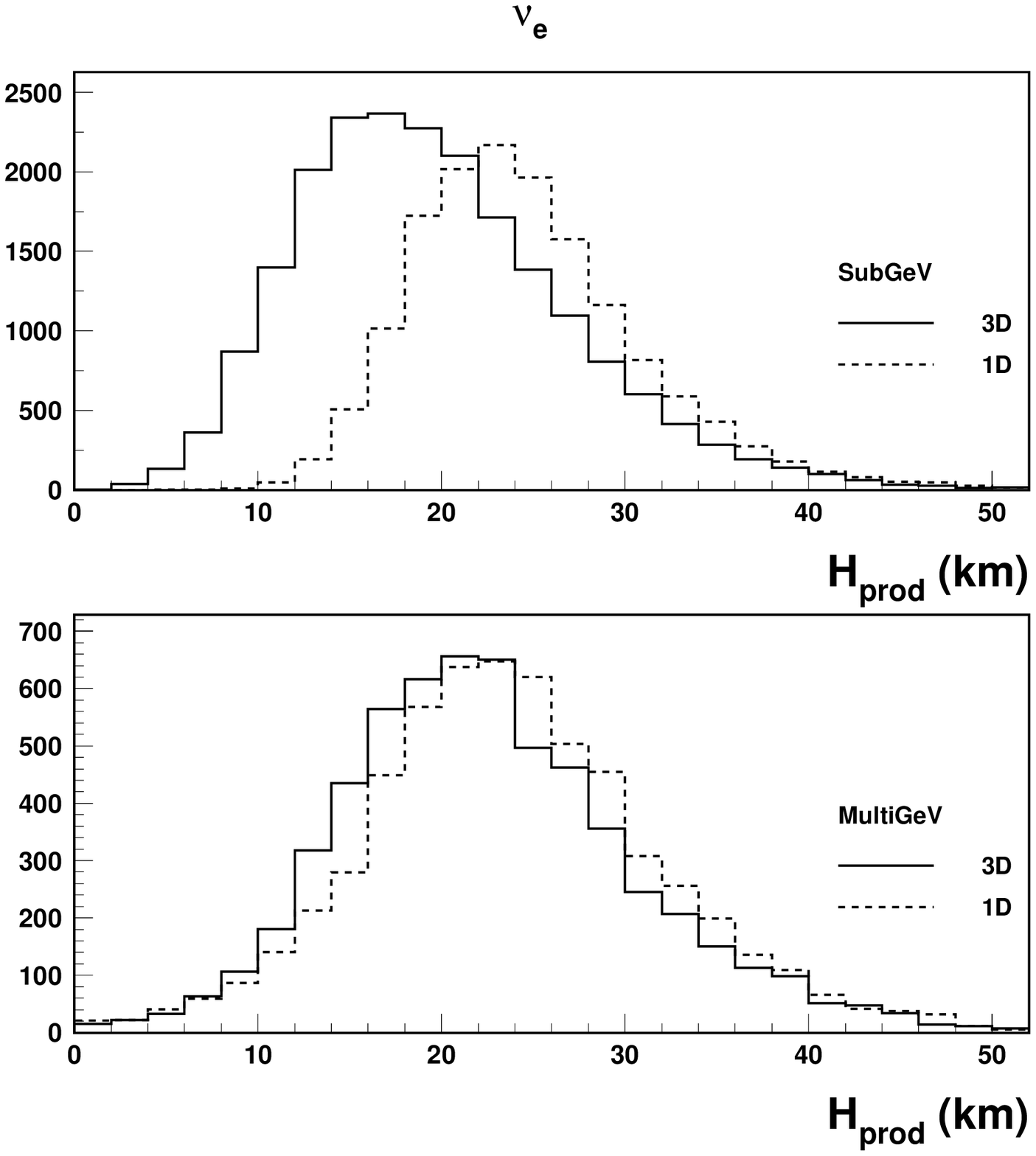,height=15.cm}} 
\caption{\em 
Distribution of production height of Sub--GeV (up) and Multi--GeV
$\nu_e$ and $\nubar_e$ around the horizontal ($|cos\theta |<0.2$) 
as obtained with FLUKA, comparing 3--D
and 1--D simulations.
\label{f:result4}}
\end{center}
\end{figure}

Other important differences are visible in the distribution of
production height of neutrinos, again only in the low energy
range, where the distribution is broader in the 3--D case than in the 1--D
case. This is particularly evident around the horizontal direction (see 
Fig. \ref{f:result3} and Fig. \ref{f:result4}).
 
\begin{figure}[htbp]
\begin{center}
\mbox{\epsfig{file=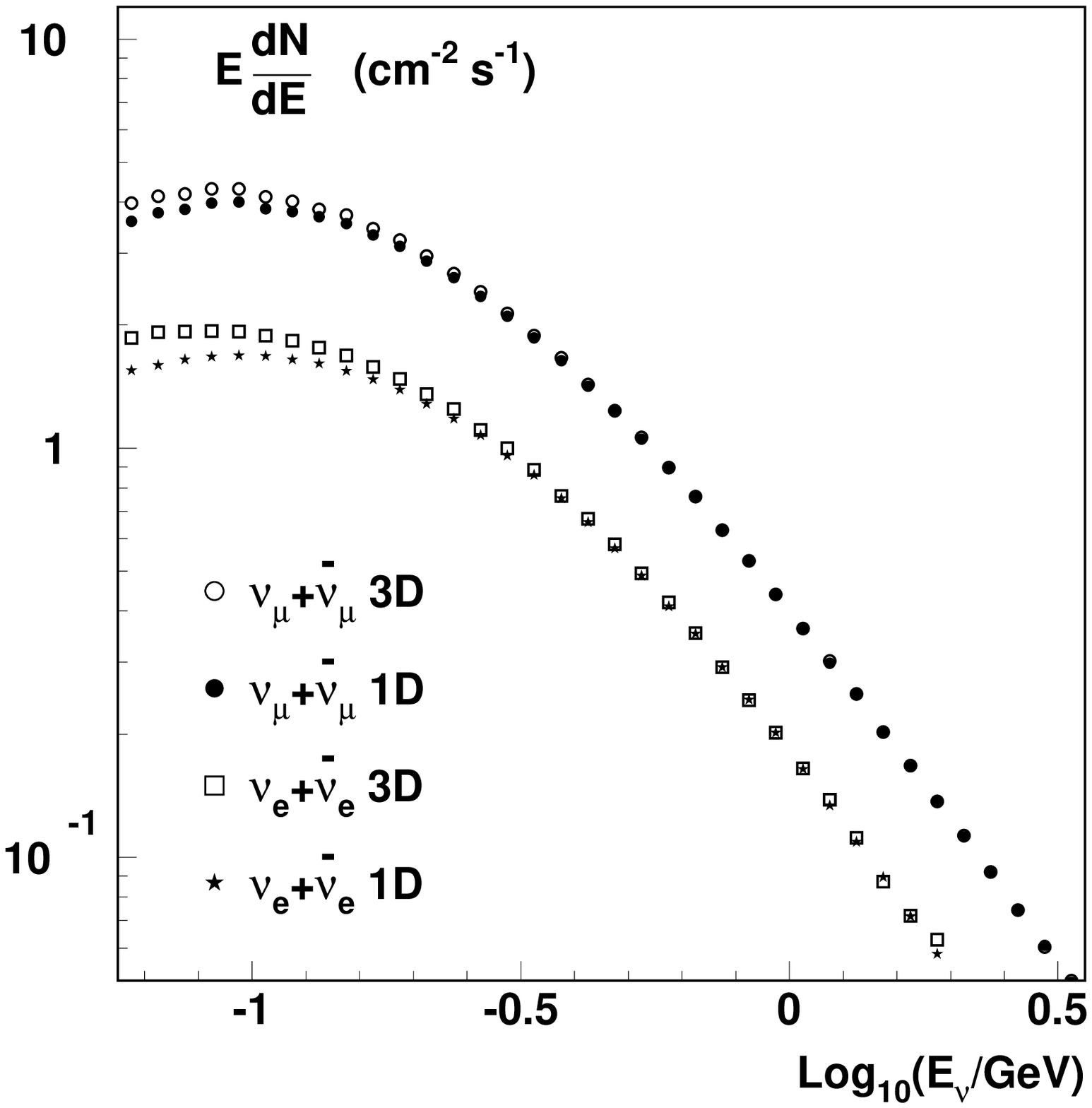,height=15.cm}}
\caption{\em Comparison of angular integrated flux for different
$\nu$ flavors 
for the 3-Dimensional and the collinear (1-D) approach,
as obtained in the FLUKA simulation with spherical geometry, 
for a generic point on the Earth surface (no 
geomagnetic cutoff is applied). \label{f:spectrum}}
\end{center}
\end{figure}

As far as the normalization of the integrated flux is concerned, a small excess 
is obtained at low energy, as shown in Fig. \ref{f:spectrum}.
 
\subsection{Detected Neutrinos and Reconstructed Direction}
 
We can distinguish between the flux of neutrinos, 
and its angular distribution 
on one side, and, on the other, 
the rate of neutrino interactions (typically of the charged current
type) with their reconstructed direction distribution.
Since we have already
stated that the Sub--GeV neutrinos as the critical ones, we shall restrict 
to this range the discussion. In this energy range, nuclear effects play an
important role in modifying the kinematics of neutrino interactions with
respect to the free nucleon case. The impact on the event reconstruction is
not at all negligible and depends on the detector characteristics.
The rate of low energy $\nu_\mu$ and $\nu_e$ CC interactions
is largely dominated by quasi--elastic scattering. The outgoing lepton
direction is often assumed to be  a good approximation of
  the incident neutrino 
direction. 
It is fundamental to consider the neutrino interaction in the nuclear
environment.
Modifications of the $Q^2$ distributions are expected 
due to Pauli blocking and threshold effects. 
In quasi-elastic scattering, the average $Q^2$ for 
a 1 GeV $\nu_\mu$ ($\nubar_\mu$) is about $0.4$ ($0.2$) GeV$^2$/c$^2$;
at 300 MeV these average values drop to $\sim0.10$ ($0.06$) GeV$^2$/c$^2$.
At these energies and above, the approximation of independent nucleons
in a potential well can be used, since 
$Q^2/2m \aprge 0.030$ GeV, the typical kinetic energy of 
a bound nucleon.
The Fermi
momentum of the struck nucleon adds a smearing to the lepton angular
distribution and to the momentum of the lepton-nucleon system, that is
balanced by the momentum of the nuclear recoil. 
 Moreover, the struck nucleon can re--interact in the nuclear
medium, with changes in the energy, number and identity of reaction
products.   The 
residual nucleus is generally left in an excited state, due to the removal
of deeply bound nucleons and to the differences in binding energy: this
excitation energy is dissipated through particle evaporation and gamma
ray emission. 
The overall momentum and energy are of course conserved, but they can be
exactly 
reconstructed only by detecting all reaction products, including the recoil
nucleus, all the low energy evaporation products (often
neutrons) and gamma rays from nuclear de--excitation.
The inaccuracies in the event reconstruction lead to a smearing of the 
reconstructed neutrino angular distribution, that could partially obscure 
the differences between the 3-D and 1-D case. To evaluate the effect under 
several ``experimental'' conditions, charged current neutrino interactions
in Argon have been
simulated  starting from the 3-D and 1-D calculated flux. 
This was
done in the framework of the same modelling tool, since
the FLUKA MC code now includes the description of neutrino--nucleus
interactions \cite{notaica,cross}, fully embedded in the FLUKA nuclear
interaction model. For quasi--elastic interactions,
the model of  \cite{lewell} for $\nu$--N interactions has been 
followed. Details of this implementation and of the FLUKA code can
be found elsewhere\cite{notaica,fluka,ferrari96}.  
We only remind remark here that it is a sophisticated intranuclear cascade 
plus pre--equilibrium code, which is very  successful in reproducing
low energy phenomenology in hadron-nucleus collisions\cite{fluka}. 

\begin{figure}[htbp]
\begin{center}
\mbox{\epsfig{file=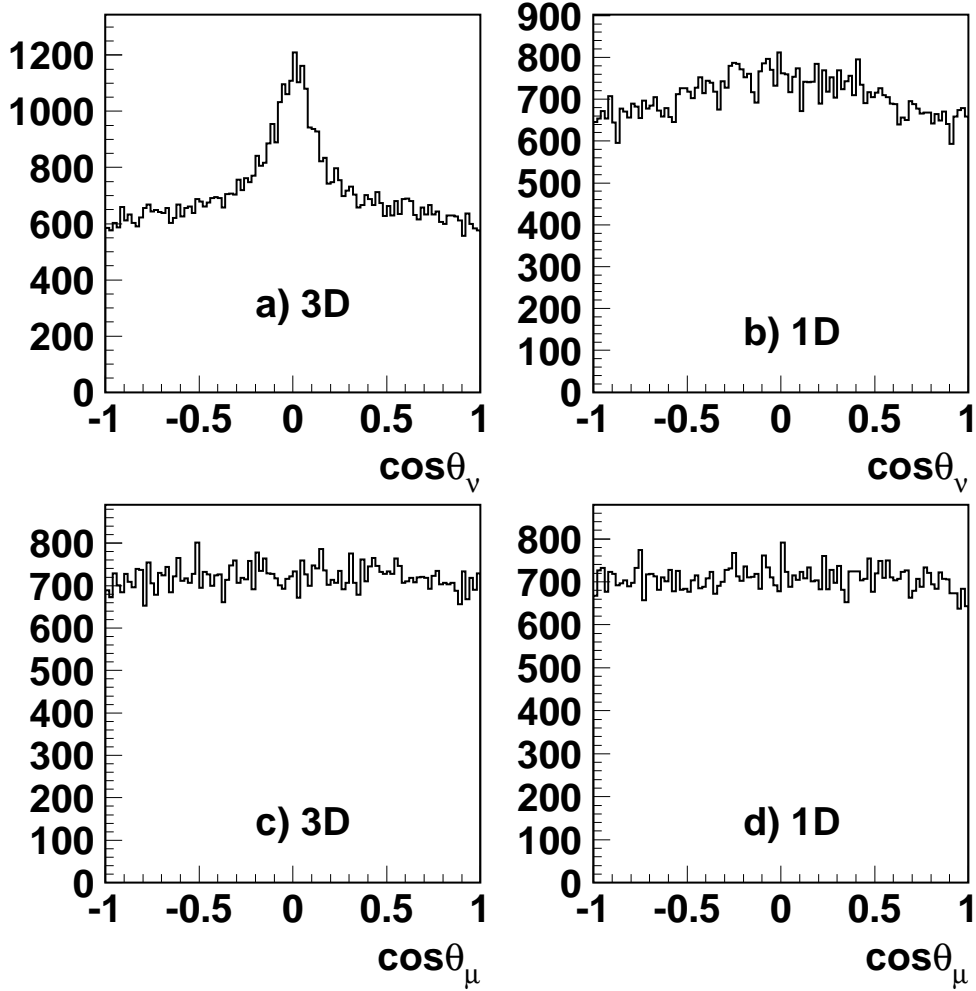,height=15.cm}}
\caption{\em Zenith angular distribution for CC--interacted $\nu_\mu$ 
events (E$_\mu<$ 1.33 GeV)
in four cases:
a) true neutrino angle (3--D simulation)
b) true neutrino angle (1--D simulation)
c) using only lepton reconstruction (3--D simulation)
d) using only lepton reconstruction (1--D simulation)
The statistical sample corresponds to an exposure of 1000 kton$\cdot$yr.
No geomagnetic cutoff is applied and no detector resolution is considered.
\label{f:numu1}}
\end{center}
\end{figure}

The simulated events have been used to reconstruct the neutrino angular
 distribution, under different assumptions. Let us start 
considering the case in which only the
outgoing lepton is measured, as in SuperKamiokande selection of single ring events.
Fig. \ref{f:numu1} shows the angular distribution of Sub--GeV muon 
neutrinos undergoing
CC interactions, when no flavor
oscillations are considered, in the 3--D (a) and 1--D (b) approaches.
The simulated sample corresponds to an exposure of 1000 kton$\cdot$yr.
No detector resolution has been considered and we remind that no geomagnetic
cutoff has been yet introduced.
In (c) and (d) we show the reconstructed
angular distribution when the direction of neutrinos is assumed to be that of 
detected muon. We can notice that
 the smearing of the $\nu-\mu$ angle is large
enough to wash out almost all 
the differences between 3--D and 1--D simulations.
 
\begin{figure}[htbp]
\begin{center}
\mbox{\epsfig{file=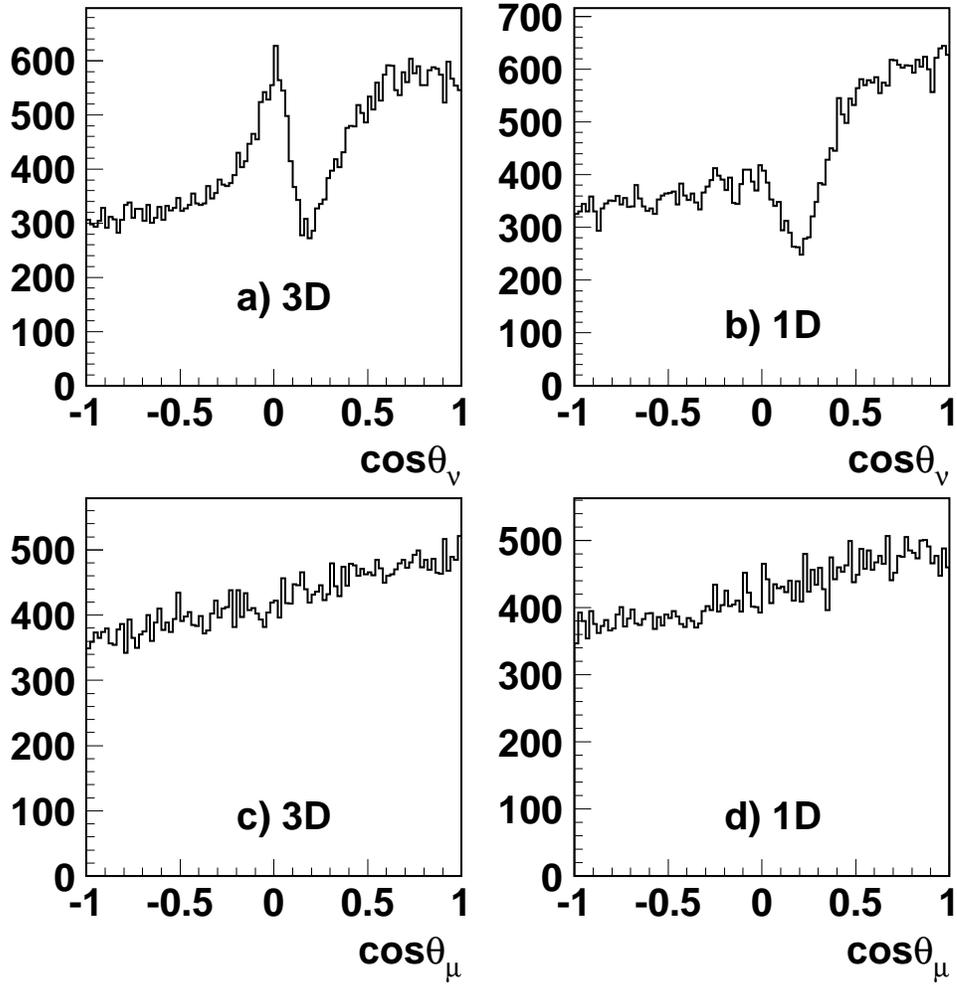,height=15.cm}}
\caption{\em Same as fig. \protect\ref{f:numu1}, when
maximal mixing $\nu_\mu$ oscillation is introduced, for
$\Delta m^2$ = 5$\cdot $ 10$^{-3}$ eV$^2$. 
a) true neutrino angle (3--D simulation)
b) true neutrino angle (1--D simulation)
c) using only lepton reconstruction (3--D simulation)
d) using only lepton reconstruction (1--D simulation)
\label{f:numu2}}
\end{center}
\end{figure}

This is still valid even in presence of factors, like flavor oscillations,
 distorting the neutrino angle distribution. This is visible in
 Fig. \ref{f:numu2} where the 4 cases of Fig. \ref{f:numu1} are shown
 for a maximal mixing $\nu_\mu$ disappearance due to flavor oscillation 
with $\Delta m^2 =5\cdot 10^{-3} eV^2$
As expected, when instead Multi-GeV events are selected,
no real difference exists between the
3--D and 1--D predictions even when the true neutrino angle is
considered.
 
The additional information carried by charged particles produced
in neutrino scattering, which 
can be identified in detectors with higher resolution
than SuperKamiokande,  has been considered as a tool to
improve the direction resolution \cite{highres}. However, 
the smearing 
introduced by the
nuclear effects 
results to  be  still large enough to obscure the 3--D  effects.
We have also investigated the possible additional benefits
coming from the detection of low energy evaporation products
(mainly neutrons), although
these are expected to be isotropically distributed.
 The only way to recover the correct distribution 
would be to measure the 
momentum of the recoiling nucleus, which is  unfortunately not possible, 
even in very high resolution detectors. 

\begin{figure}[htbp]
\begin{center}
\mbox{\epsfig{file=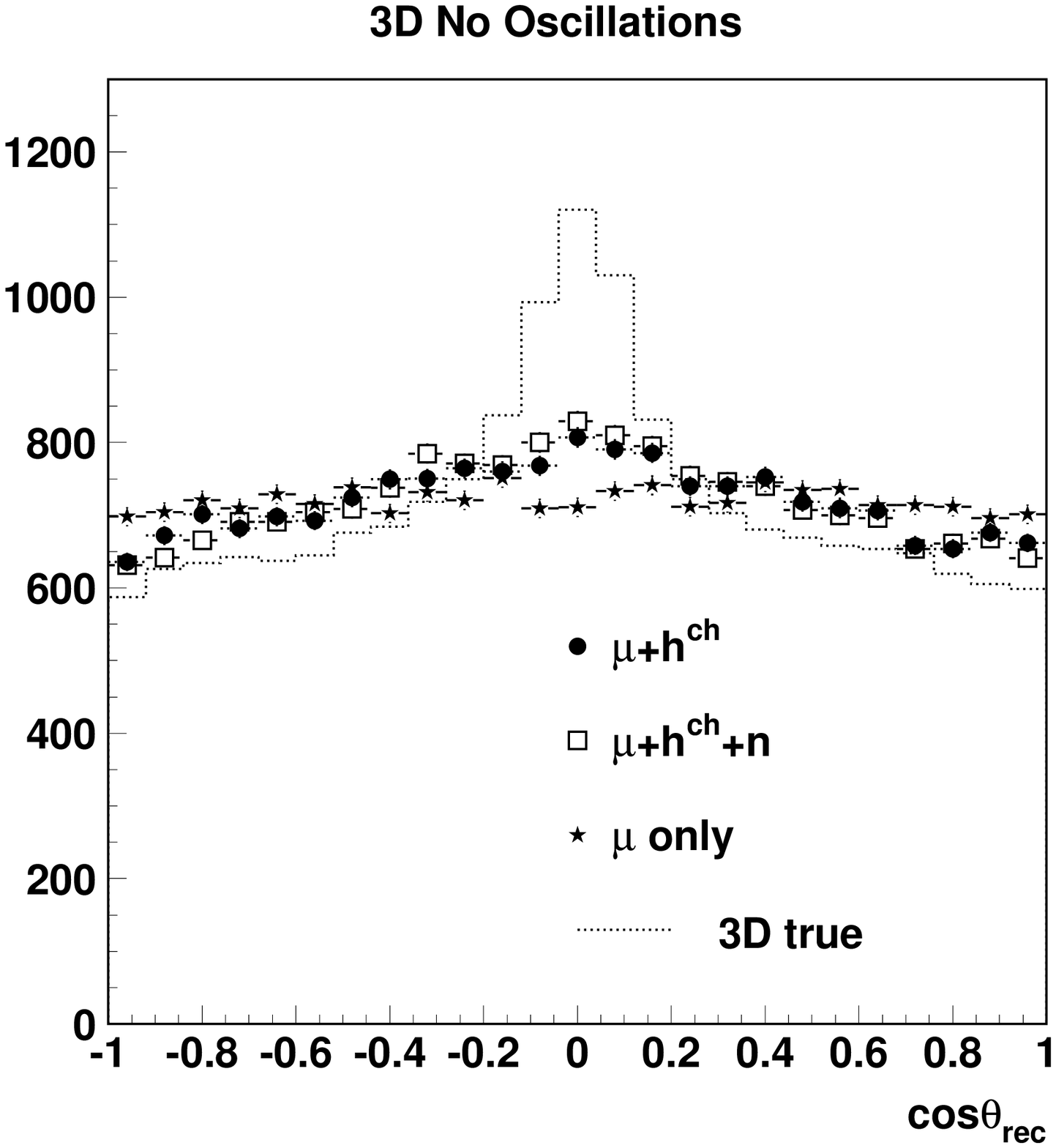,height=15.cm}} 
\caption{\em
Reconstructed zenith angular distribution for CC $\nu_\mu+\nubar_\mu$ interactions
(E$_\mu<$ 1.33 GeV, 1000 kton$\cdot$yr exposure) in three cases:
a) when the direction reconstruction is given by the combination
of the reconstruction of the muon and of the possible charged hadrons;
b) when also the neutrons from nuclear de-excitation are included.
c) when only the muon is measured.
For reference, the true angular distribution are also shown.
 \label{f:recoil}}
\end{center}
\end{figure}

The results are reported in Fig. \ref{f:recoil}, where the reconstructed angular distribution
of $\nu_\mu + \nubar_\mu$ neutrinos (no oscillations) is shown for
the 3--D simulation only, comparing three different cases: 
i) when the charged lepton and hadrons are measured, ii) when also 
evaporation neutrons are added; iii) when only the lepton is 
reconstructed. 
We are still assuming the ideal case of perfect energy and direction resolution for
the detected particles. 
The true 3--D angular distribution is also shown for comparison.
Some improvement can be noticed when charged hadrons are considered.
The addition of neutrons does not give a significant improvement.
Some advantage comes
from the fast neutrons which are produced in nuclear re--scattering, but they
are very few.
 
\begin{figure}[htbp]
\begin{center}
\mbox{\epsfig{file=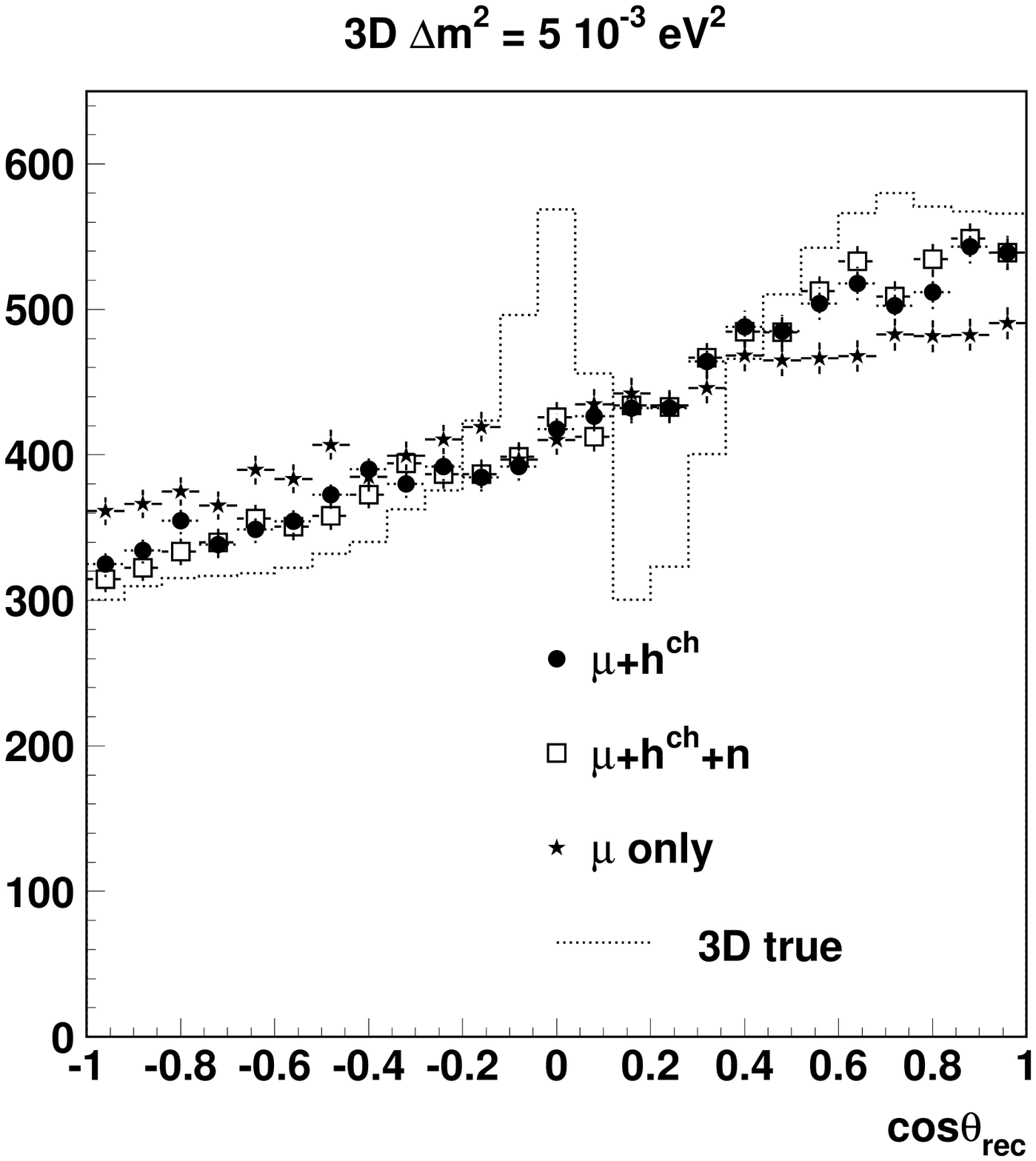,height=15.cm}} 
\caption{\em
Reconstructed zenith angular distribution for CC $\nu_\mu+\nubar_\mu$ interactions
(E$_\mu<$ 1.33 GeV, 1000 kton$\cdot$yr exposure) in the same cases
of Fig.\ref{f:recoil} for maximal mixing oscillations with
$\Delta m^2$ = 5$\cdot$10$^{-3}$ eV$^2$.
 \label{f:recoil2}}
\end{center}
\end{figure}

The same cases of Fig.\ref{f:recoil} are repeated in Fig.\ref{f:recoil2}
when maximal mixing $\nu_\mu$ oscillations are introduced 
with $\Delta m^2$ = 5$\cdot$10$^{-3}$ eV$^2$.

\begin{figure}[htbp]
\begin{center}
\mbox{\epsfig{file=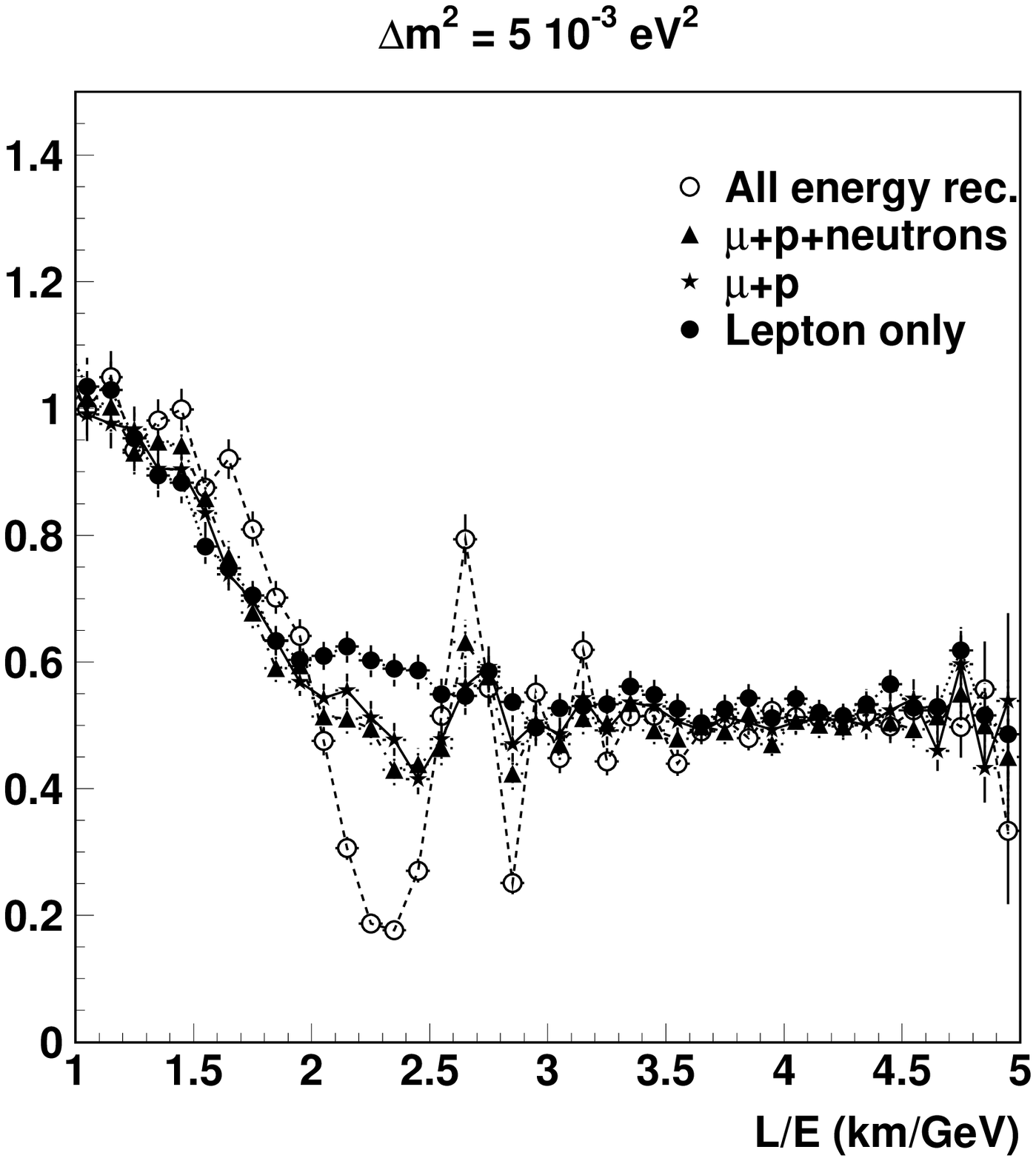,height=15.cm}} 
\caption{\em
Ratio
of the L/E distribution of simulated $\nu_\mu + \nubar_\mu$
events in the case of maximal mixing oscillations and 
$\Delta m^2$ = 5$\cdot$10$^{-3}$ eV$^2$
to that of the same sample when no oscillations are present, for different event
reconstructions, CC $\nu$ interactions are selected.
\label{lsue}}
\end{center}
\end{figure}

As a final exercise on this topic, we looked at the  
muon neutrino oscillation pattern,
 following the analysis method described
in \cite{ppicchi,batlip}.  The number of events is plotted 
as function of L/E, where L is the
estimated path of neutrino from production to detection and E is the 
neutrino energy. In Fig.\ref{lsue}
we show the ratio of the L/E distribution of simulated $\nu_\mu +
\nubar_\mu$ events in the case of maximal mixing oscillations and $\Delta
m^2$ = 5$\cdot$10$^{-3}$ eV$^2$ to that of the same sample when no
oscillation are present, for different event reconstructions.
Once again the simulated statistics corresponds to an exposure of
1000 kton$\cdot$yr (3--D simulation) and
no experimental
resolution is applied.
The statements concerning the importance of the recoil of the nucleus
are confirmed: the oscillation pattern is totally hindered when  the
lepton alone is detected, and is strongly suppressed unless the  event is
fully reconstructed. Of course, if high energy events are selected (typically
when $P_\mu>$ 1$\div$2 GeV), the oscillation pattern can be recovered
even with the muon reconstruction alone.

\section{Inclusion of Geomagnetic Cutoff}
 
\begin{figure}[htbp]
\begin{center}
\mbox{\epsfig{file=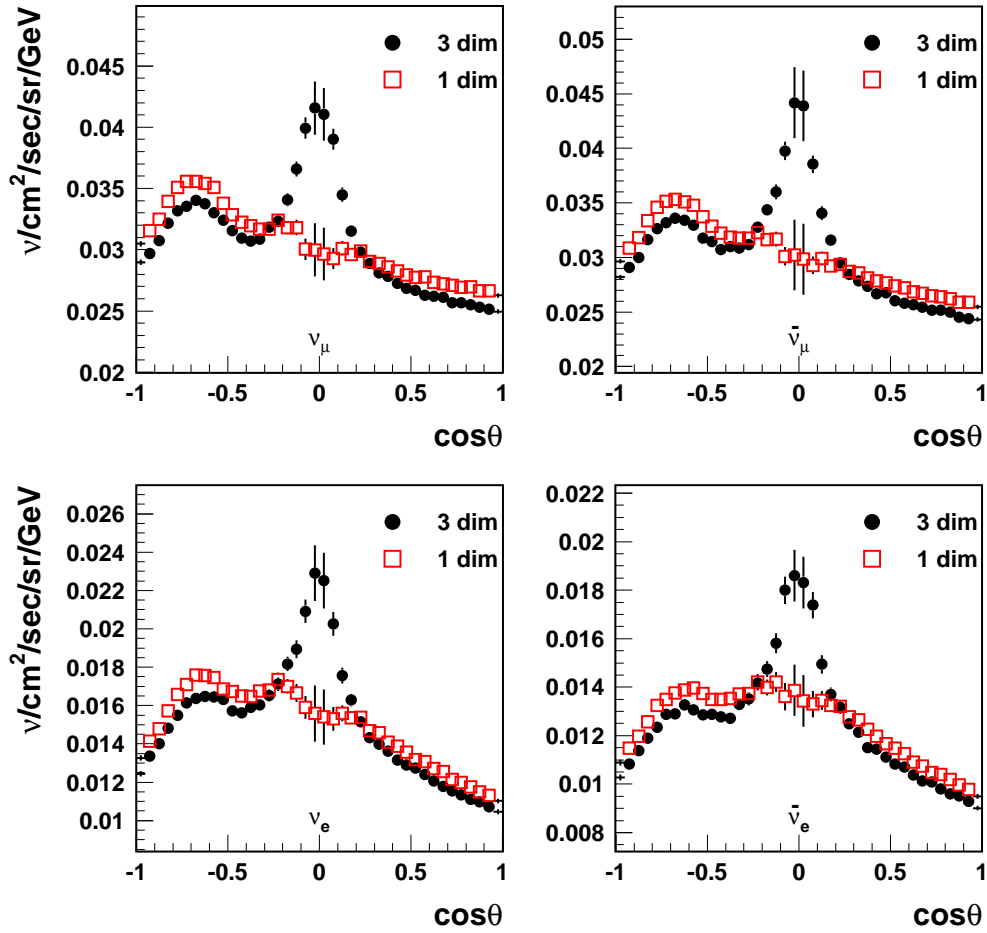,height=15.cm}}
\caption{\em Neutrino flux as a function of Zenith angle, for the different
flavors, as predicted for the site of SuperKamiokande. Results from
3--D and 1--D simulations are compared. \label{fluxcth}}
\end{center}
\end{figure}

Further differences between the 3--D and 1--D calculations could be expected
in principle when geomagnetic cutoffs are applied to the primary nucleons and 
nuclei.
Technically, this is usually accomplished applying a rejection algorithm
to the unperturbed flux. It depends on the geomagnetic location of the detector,
on the geomagnetic location and altitude of primary interaction, primary
rigidity and on the zenith and azimuth angles of the arrival direction of 
primary.
In the collinear approximation, the coordinates and arrival direction of primary
are instead strictly linked to those of the neutrino. 
We might expect, in principle, a different result in the
energy-angle correlation, again for Sub--GeV neutrinos. In order to check this, 
we introduced the geomagnetic cutoff in our simulation runs, both
3--D and 1--D, by means of the technique of anti--proton back--tracing, 
expressing the geomagnetic field
as an expansion in spherical harmonics, similar to
that described in the work of Honda et al. \cite{honda}. This has been done
for different geographical sites.
Similar results have been obtained with a faster algorithm, in which
the geomagnetic field is described by a dipole properly off-centered with 
respect
to the Earth sphere \cite{roesler}.
As an example, in Fig. \ref{fluxcth} we show the 
neutrino flux at low energy for different flavors as a function of $cos\theta$
in the site of SuperKamiokande, comparing the 3--D and 1--D results.
 
\begin{figure}[htbp]
\begin{center}
\begin{tabular}{c}
\mbox{\epsfig{file=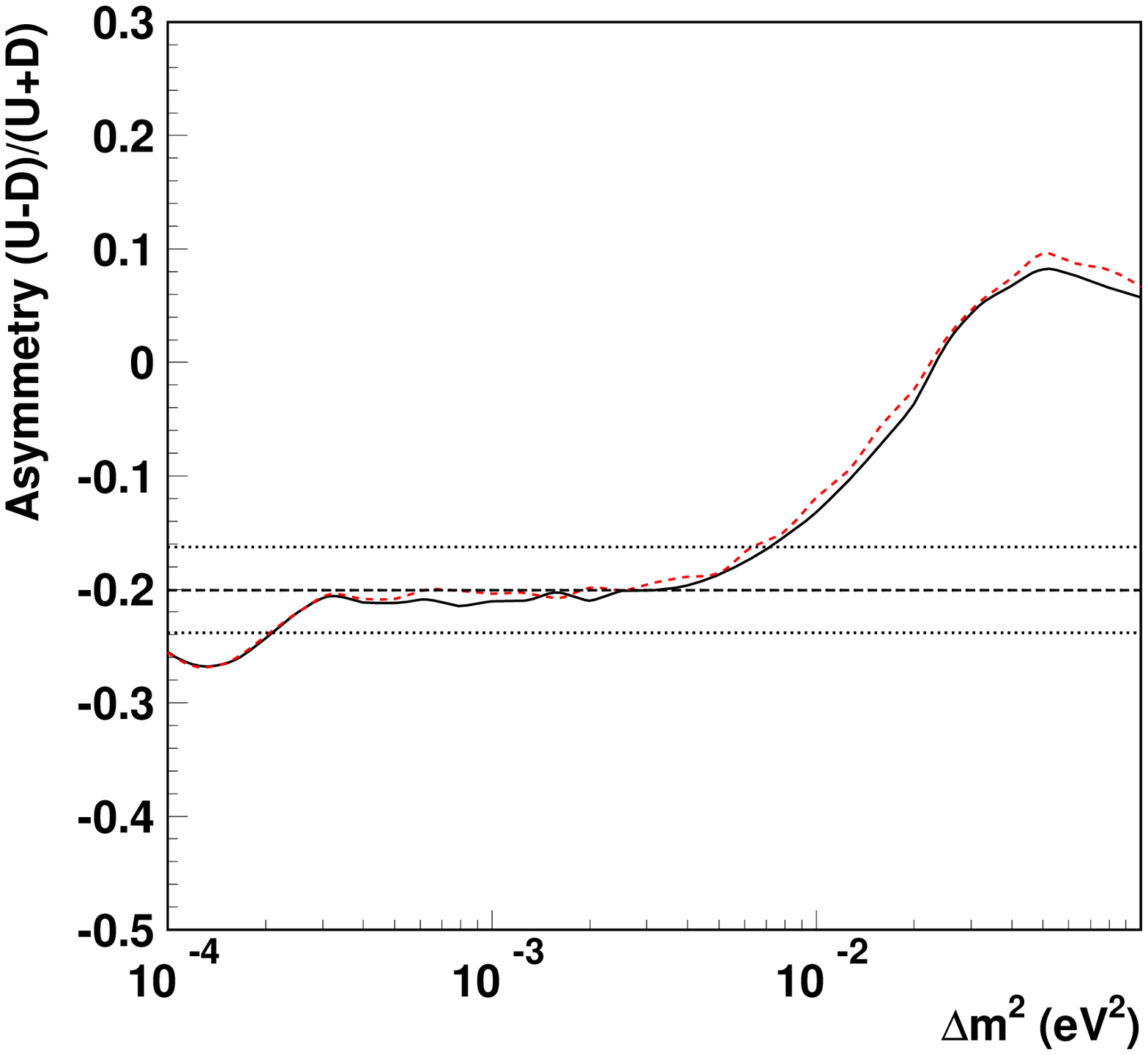,height=8.cm}} \\
\mbox{\epsfig{file=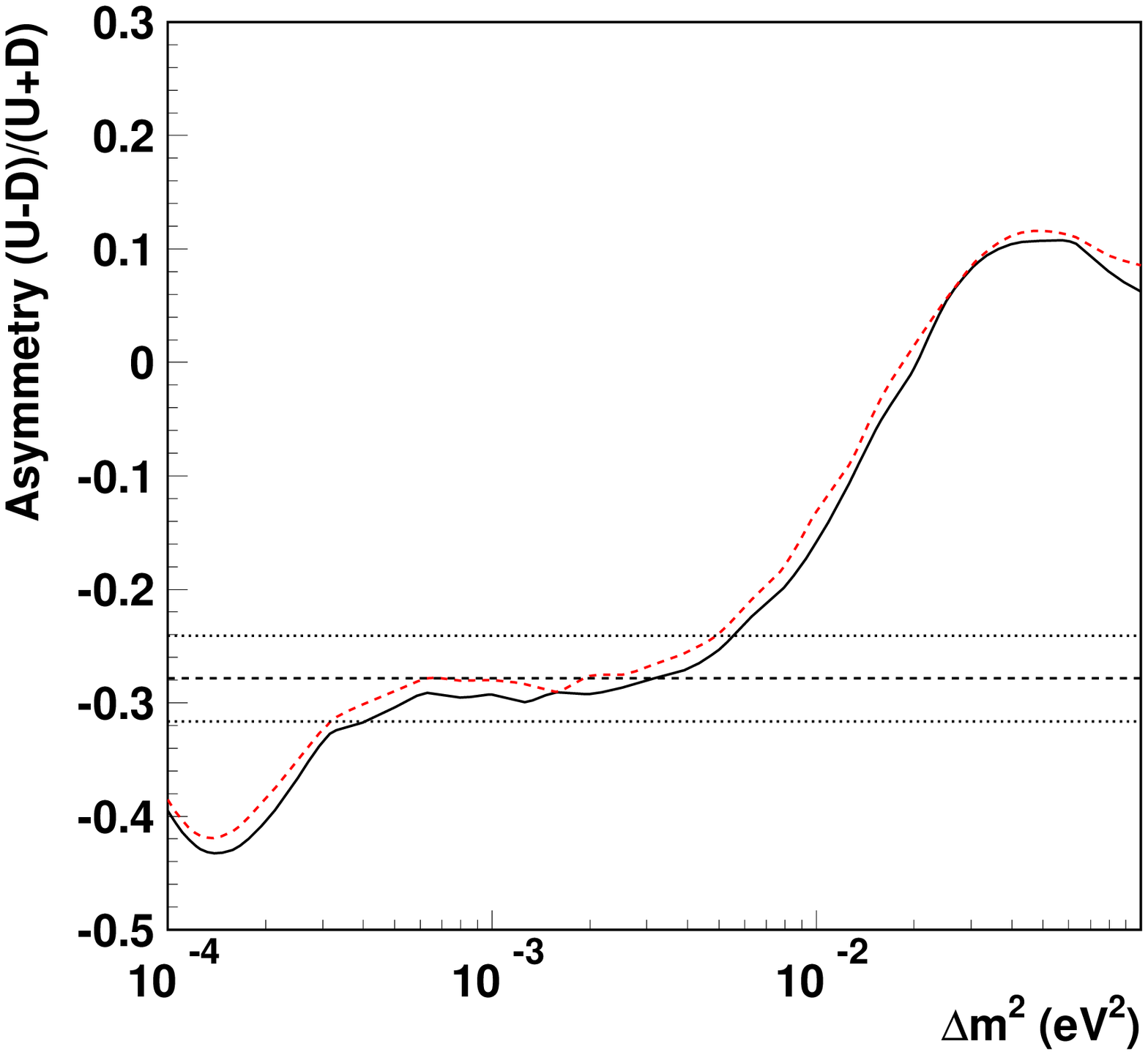,height=8.cm}} 
\end{tabular}
\caption{\em Simulation of the asymmetry parameter for $\nu_\mu$+$\nubar_\mu$ 
in case of maximal mixing oscillation, as a function of $\Delta m^2$, for
the Sub-GeV events in SuperKamiokande. Upper panel: using reconstructed
neutrino direction for single ring events. Lower panel: the same result
assuming that the true neutrino direction is know.
The continuous (dotted) line is the prediction from the 3--D (1--D) simulation.
The band relative to the total uncertainty of SuperKamiokande after
45 kton$\cdot $yr is also shown. \label{f:cutoff1}}
\end{center}
\end{figure}

The up--down differences are now visible. The enhancement around
the horizontal direction for the 3--D simulation remains visible,
but it will be smeared after neutrino detection.
In order to show this, we have calculated the up--down asymmetry parameter
for Sub-GeV muon neutrinos under the hypothesis of a maximal mixing oscillation into
an undetected flavour, as a function of $\Delta m^2$.
In order to make a realistic exercise, we have introduced 
selections similar to those adopted in the recent 
SuperKamiokande analyses \cite{superknew}, {\it i.e.} 
using the muon direction, excluding
a region around the horizontal ($|cos\theta | \leq$ 0.2) 
and accepting lepton momenta above 0.4 GeV/c.
The result is shown in
in the upper part of Fig. \ref{f:cutoff1}. 
The 3--D and 1--D approaches are identified by different 
line styles. 
The expected uncertainty region of SuperKamiokande after the analysis
of the 45 kton exposure sample is reported \cite{superknew}
(we considered the statistical error and the quoted
5\% systematic error added in quadrature).
Differences exist, but they are again small. 
Once again, the smearing in the reconstructed angle
introduced by neutrino interactions obscures the effect.
In the bottom part of the 
same figure we show what would be obtained if the true neutrino direction
could be recovered. Even in this hypothetical case, the differences 
comparing 3--D are 1--D
are too small. 
 
\begin{figure}[htbp]
\begin{center}
\mbox{\epsfig{file=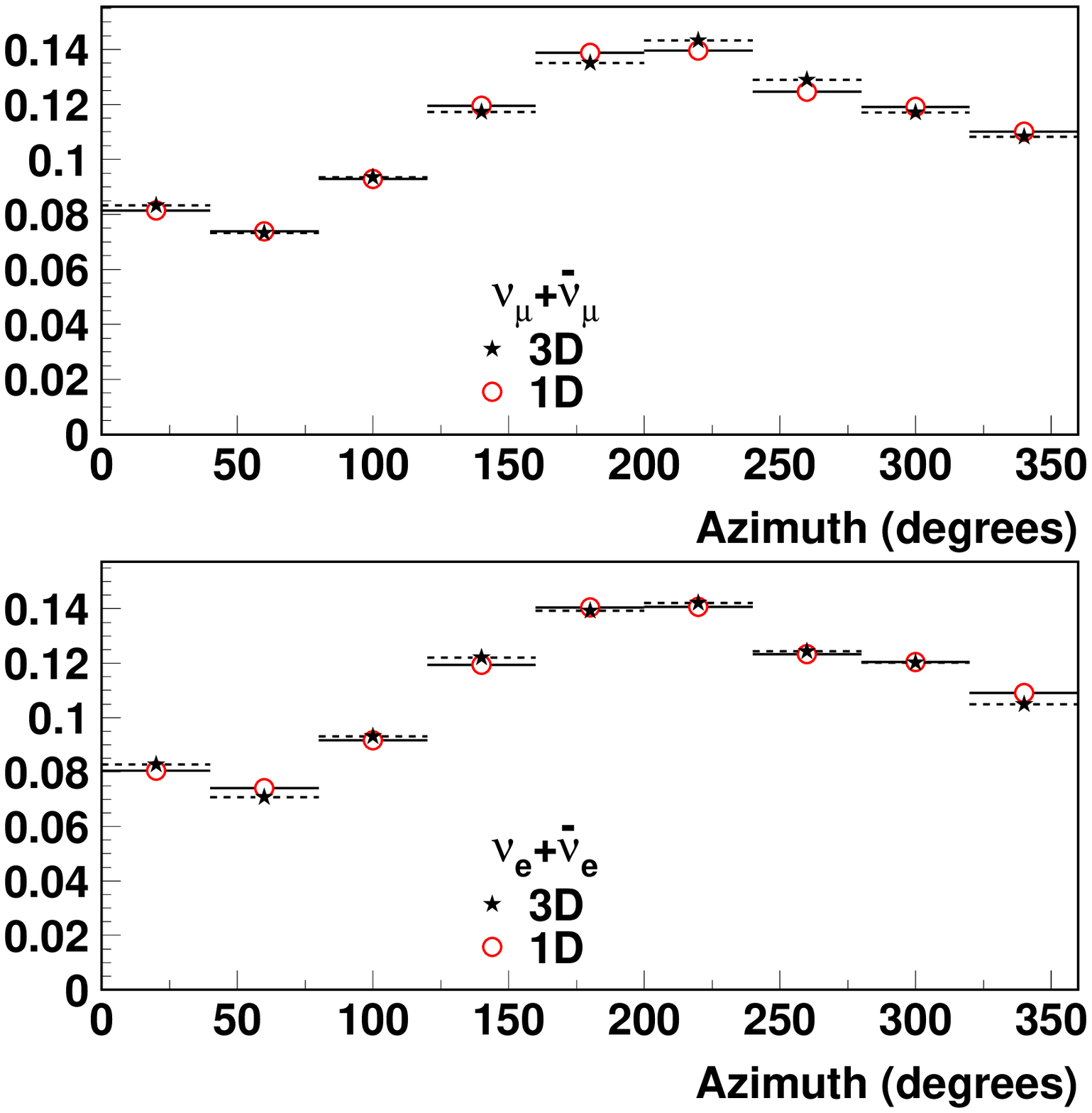,height=15.cm}} 
\caption{\em Simulation of the east-west effect (without oscillation) for
the SuperKamiokande site for the two neutrino flavors.
The 3--D and 1--D simulation in the spherical geometry are compared.
\label{ew}}
\end{center}
\end{figure}

The East-West effect has been also considered, but again
no noticeable difference has been detected between the 3--D and 1--D approaches,
as shown in Fig. \ref{ew}, where the situation of SuperKamiokande (without
oscillations)
has been simulated, using the same selection cuts described in \cite{eastwest}.

A posteriori, these results are not surprising. As we have shown
in the plots when no cutoff was yet introduced, the 3--D effects
introduce an enhancement around the horizontal direction, but with 
a perfect up--down symmetry. The small differences that can be 
noticed in Figg. \ref{f:cutoff1} and \ref{ew} are essentially due to
oscillation effects coming from
differences in the path lengths of neutrinos.
 
\section{Conclusions} 
 
We have shown that the collinear approximation
that has been used for many evaluations of atmospheric neutrino fluxes
is based upon a wrong assumption and leads to an incorrect angular distribution
of neutrinos at the Earth's surface. This turns out to be relevant for
Sub-GeV energy range.
However, a precise  experimental momentum reconstruction is hardly achievable
due to the 
the features of neutrino-nucleus charged current interaction at low 
energies. In realistic situations, this induces a the smearing on the
measured angular distribution that is such to obscure in great
part the differences with respect to a full 3--D simulation.
This is true not only when just the outgoing lepton is measured, as in the 
single ring
events of SuperKamiokande, but also when the recoiling nucleon is measured as 
well, provided that the residual recoiling nucleus is undetected.
In summary, the consequences for physics analyses following from the use of
the collinear approximation in the flux calculation are very small.
In particular, referring to the question of
neutrino flavor oscillations,
the limits on $\Delta m^2$ affected by the angular distribution of
Sub-GeV neutrinos, should remain practically unchanged.
Nevertheless, experimental analysis aiming to achieve the maximum possible 
precision should make use of a full 3--D simulation in spherical geometry.
There are other experimental observables which could be more directly affected
by the 3--D features. For example, the measurement
of the low energy muon flux in the high atmosphere is often considered 
an important benchmark for the Monte Carlo codes for atmospheric
neutrinos. We expect that the agreement of existing calculations with
the present data might be improved considering the 3--D simulation, and in this 
case
there is no intermediate smearing as that emerging from neutrino--nucleus
interaction.
Again, we expect the differences to show up preferentially around
the horizontal direction, but the existing measurements were all performed
around the vertical. Possibly, new experiments could consider the possibility
of increasing the angular acceptance of the detector.
 
\section*{Acknowledgments}
We wish to thank Prof. C. Rubbia, the Icarus collaboration
and the MACRO collaboration for the strong support
to this effort. We also acknowledge the stimulating discussions
with T.K. Gaisser and T. Stanev, who also provided us with the primary spectrum
used by the Bartol group.

\newpage

\section*{Appendix: An attempt of analytical description}
 
The assumption at the basis of the collinear approximation 
is just that the effects of $\theta_{N\nu}$ are cancelled once
that the primary particles are isotropically distributed.
Such hypothesis is in principle wrong, and
before entering in the examination of the calculation results,
 it can be instructive
to consider a simple case, that can be analytically described, which
demonstrates that.
 
It is the case of a planar geometry, in which an  isotropic flux 
of primary particles $\phi_0$ fills  half the
space (the  region $z < 0$). 
The plane  $z=0$ acts as a perfect absorber:
all  primary particles that  touch the plane
are  absorbed and during this process each one of these particle  emits
an average number of neutrinos  $\langle n_\nu\rangle$.
The  capture rate $C$  (the number of primary particle
captures per unit time  and  unit surface) is  easily calculated
as: 
\begin{equation}
 C = 2 \pi\, \int_0^1 dx~ x \,\phi_0  =  \pi\; \phi_0 
\end{equation}
where we have assumed  that all primary particle  are  ultrarelativistic,
set  the speed of light to $c=1$,  and defined
$x = \cos \theta_0$ as the component  of the velocity
of the primary particle  orthogonal to
the absorbing  plane ($x=1$  corresponds to a particle moving  
orthogonally and toward the  plane).
Obviously the rate of  absorption   is linear in $x$.
The absorbing plane is  a source of neutrinos.
Integrating over all  emission   angles  
the  number of neutrinos emitted  per unit  time  and unit  of surface
is  obviously:
\begin{equation}
S_\nu = C~ \langle n_\nu \rangle = \pi \,\phi_0 \,\langle n_\nu\rangle.
\end{equation}
We are  interested  in  the  angular  distribution 
of the emitted   neutrinos $dS_\nu/dy$   where  $y = \cos \theta_\nu$ is
the component of  neutrino direction  orthogonal to the
absorbing plane.
We can write in  general:
\begin{equation}
 {dS_\nu \over dy}  =  \pi\; \phi_0  \langle n_\nu\rangle \; f(y)
\end{equation}
where  the function $f(y)$ satisfies  the normalization
condition
\begin{equation}
 \int_{-1}^{+1} dy ~f(y) = 1.
\end{equation}
 
To compute  the  form of   $f(y)$ we need to specify 
the angular distribution  of the neutrinos with respect
to the primary particle direction.  Two  limiting cases  are  
easy to obtain. 
In the case where  the neutrinos  are emitted 
isotropically  we have 
 $f_{iso} (y) = {1\over 2}$.
In the case  where  the neutrinos  are  exactly collinear
to  the primary particle  direction
we have $f_{coll} (y) = 2 y~\theta(y)$, where the  function  $\theta$ 
is  defined as: $\theta(x) = 0$  for  $ x < 0$ 
and $\theta (x) = 1$  for  $ x \ge  0$.
 
In order to solve 
the most  general  case of  an arbitrary angular distribution
we  can solve the special  problem where the neutrinos  is  emitted
at a fixed angle  $\alpha$  with respect to the primary direction
(with an  additional azimuthal  angle $\beta$  uniformly  distributed
between  0  and $2 \pi$.  The case of an  arbitrary distribution
of  primary--neutrino angle 
can then be obtained  with a  simple integration.
 
The  source of neutrinos in  a given  direction 
can  be obtained   summing  over  all  appropriate  combinations
of  primary direction and  azimuthal   emission  angle:
\begin{equation}
 f_\alpha (y) = 
~2\,\int_0^1 dx ~x ~\int_0^{2 \pi} {d\beta \over 2 \pi} 
~\delta [y - (x\,\cos\alpha - \sqrt{1-x^2} \,\sin \alpha\, \cos \beta) ]
\end{equation}
The  result  of the integration is:
\begin{equation}
f_\alpha(y) = \cases { 
0 ,       &for $y < - \sin \alpha$ and $\cos\alpha > 0$ \cr
\omit      & or  $y > \sin \alpha$ and $\cos\alpha < 0$, \cr
2\, \cos \alpha \,y,  &for $y  > \sin \alpha$ and $\cos\alpha > 0$ \cr
\omit                 & or  $y < -\sin \alpha$ and $\cos\alpha < 0$, \cr
F_\alpha (y)  & otherwise. }
\end{equation} 
with
\begin{equation}
 F_\alpha (y) = 
\cos \alpha \; y + {2 \over \pi} \sqrt{1 - \cos^2 \alpha - y^2} 
- {2 \over \pi} \cos \alpha \; y ~ \tan^{-1} \left [
{ \cos \alpha \; y \; \sqrt{1 - \cos^2 \alpha - y^2}  \over 
\cos^2 \alpha + y^2 -1 } \right ]
\end{equation}
 
The resulting angular distribution is therefore different
from the collinear expectation. This is recovered
in the limit  of $\alpha \to 0$, where we have
$f_\alpha (y)  \to 2 y \; \theta (y)$.
Averaging  over all emissions  angles  we  also  find:
\begin {equation}
\int_{-1}^{+1}  d\cos \alpha ~f_\alpha  =  {1 \over 2}
\end{equation}
as  expected for an isotropic distribution.
Note that in general  we  can  distinguish three  different  angular  regions
for the emission of the neutrinos. 
In the case  $\alpha < {\pi \over 2}$  we have that
no neutrinos are  emitted in  the most
backward   directions  ($y < - \sin \alpha$).
For  the most forward directions  $y > \sin\alpha$
the  angular distribution has the 
same  linear form of  the collinear  case
but  the  normalization is suppressed : $f_\alpha (y) = \cos\alpha \ y$.
In the intermediate   region  $  -\sin\alpha \le y \le \sin\alpha$
one has  a more complex  behaviour.
The case $\alpha \ge {\pi\over 2}$ can be obtained from the previous case 
with a  reflection, that is  with the   substitutions
$\cos \alpha \to -\cos \alpha$, $y \to -y$.
 
These analytical consideration can be extended to the spherical geometry,
where 
we have a  spherical  absorbing surface  of unit  radius.
An isotropic flux  of primary  particles $\phi_0$ fills the  entire space
{\em outside} the surface.  All primary particles   that 
`touch' the spherical  surface
are  absorbed, each emitting  an average number of neutrinos
$\langle  n_\nu \rangle$.
Let  us consider  now  an observer  located  at a radius
$r$   (because of the spherical symmetry of the problem
we do  not  need  to consider the angular  position of  the observer)
 and calculate the  flux of neutrino
measured by such an observer.  
Note that  the center of  the sphere and  the observer define
a privileged  direction that breaks the 
initial spherical symmetry of the problem, and only a
symmetry  for  rotation  around this special  axis  survives.
In this discussion we will  concentrate on the case
$r < 1$ when the observer is  {\em  inside} the spherical 
surface.
It is convenient  to define  a z axis with the  origin in the center
of the sphere and passing through the observer, and define
$z = \cos \theta_\nu$  as  the zenith angle  of an  observed  neutrino.
Note  that  here  following  the 
standard definition   we define  the
zenith angle  so that:
$z = \cos \theta_{zenith} = - p_z/|\vec{p}|$:  a particle with $z=+1$  
is  moving  downward.
Our problem  is to compute the  flux  $\phi_\nu (z, r)$ of  neutrinos
with cosine of zenith  angle  $z$  for  an observer at position $r$.
This is achieved as  follows.
The line  of sight  defined  by the observer  position  $O$  and the 
direction $z$   will intersect the neutrino source surface ($r=1$)
at a point $S$.  Let us  call $C$ the center of the sphere.
Let us define $\theta_{emission}$
the  angle  $\hat {CSO}$, that is  the angle between the  normal to the 
surface  and the line of  sight from the source point to the observer.
The cosine of this  angle  has value  $y = \sqrt{1 - r^2(1-z^2)}$.
The flux of  neutrinos   can  be computed  as:
\begin {equation}
\phi_\nu (z, r)  = 
~ \left [ {L^2 \;dz \; d\varphi \over y(z,r)} \right ]
~ \left [ { dS_\nu (y) \over dy } (y(z,r)) \right ] 
~\left [ {1 \over 2 \pi L^2 dy} \right ]
~\left [ { dy \over dz} \right ]
\label {eq:spher-dem}
\end{equation}
In this  equation $L$ is  the distance $\bar{OS}$ between the observer and the 
neutrino  source   along the line of sight considered.
The   first  term  in square parenthesis  is the area    of the source 
subtended by the solid angle    $dz \;d\varphi$ 
($\varphi$  is an azimuthal angle)  with the origin  at the observer;
note the   factor $y^{-1}$  that takes  into  account the orientation
of the surface  element  with respect to  the line of sight.
The  second term  is the  number of neutrinos emitted
per unit  time    by  a unit area  element of the surface around the
direction  considered.  The third term  is the inverse of the
area  over which the emitted  
neutrinos are spread after  a distance $L$, and  finally we  have
a  Jacobian factor  between the emission and the detected angle.
Note the  expected  cancellation of $L$ 
(for  completeness: $L = -r\, z + \sqrt{r^2 z^2 + 1 -r^2}$)
in  Eq.\ref{eq:spher-dem}.  
Simplifying  the equation  we find:
\begin {equation}
\phi_\nu (z, r)  = 
{1 \over 2 \pi} ~
~ \left [{1 \over y}  { dS_\nu (y) \over dy } \right ]_{y=y(z,r)} =
\phi_0 \langle n_\nu \rangle ~\left ( f(y) \over 2 y \right )_{y = y(z,r)}
\label {eq:spher-dem1}
\end{equation}
 
At this point, 
if one aims to a more realistic description (neutrinos produces at different
heights, realistic angular distribution, etc.) the analytical approach may 
become
more complicated and less instructive with respect to a numerical test which
can be constructed by the reader by means of a simple toy simulation.
We can anyway arrive to the following statements.
In the limit   where all
neutrinos are exactly collinear to the primary particle,
the neutrino  flux inside  the
sphere is  $\phi_\nu (z,r) = \phi_0 \langle n_\nu\rangle$, it is
isotropic and  independent on the observer position $r$  (for
any $r < 1$).
In   general, for neutrinos not collinear with primary particle, 
these  properties   are  not 
valid: the neutrino  flux
does  depend on the observer position, and it is not isotropic,
except of course for  the very special case of  an  observer located at
the  very center of the sphere.

\end{document}